\documentclass{article}
\usepackage{amsmath, amssymb, latexsym, epsfig}
\begin{document}
\centerline{\large\bf Time Correlation in Tunneling of Photons}
\begin{center}
Peter Hrask\'{o}\footnote{E-mail: peter@hrasko.com}\\
Department of Theoretical Physics,\\
Janus Pannonius
University, Ifjus\'{a}g u.\ 6, 7624 P\'{e}cs, Hungary.
\end{center}

\vspace{10mm}
 
{\bf Abstract}: I propose to consider photon tunneling as
a space-time correlation  phenomenon
between the emission and absorption of a photon on the two
sides of a barrier. Standard technics based on an appropriate counting
rate formula may then be applied to derive the tunneling time
distribution without any {\em ad hoc} definition of this quantity. 
General formulae are worked out for a potential model using
Wigner-Weisskopf method. For a homogeneous square barrier in the limit of zero
tunneling probability a vanishing tunneling time is obtained.
 
\vspace{6mm}
 
\section{Introduction}
The phenomenon of tunneling was recognized immediately after the
birth of quantum theory as one of the most striking features of
microphysics. Though tunneling probabilities are predicted
unambiguously by the theory this is not the case with the time
required for tunneling \cite{MacColl32}. Tunneling time is equal
to the time interval required to travel a certain distance when a
barrier between the endpoints is present minus the time required
to travel the distance outside the barrier. This quantity is
directly measured in an experiment "with clocks" i.e. using
two detectors that signalize the moment of
departure and the moment of arrival of the tunneling particle (see Fig.1), but no
experiment of this type has so far been actually performed. No
theory is required to {\em infer} the value of the tunneling time
from such an experiment because it is just this experiment which
defines the notion "tunneling time"
operationally. It may of course happen that an
experiment if performed would give no sharp tunneling time but
rather a tunneling time distribution.
 
Theory is needed to {\em predict} the result of the
experiment. However, up to now no method has been proposed which
would
permit us to infer from quantum theory the expected value (or
distribution) of a tunneling time experiment as described above.
The Wigner-time \cite{Wigner55}, the
B\"{u}ttiker-Landauer
time \cite{Buttiker82}, \cite{Landauer89} and the
Larmor-time \cite{Baz64} are
each based on different secondary
criteria for the time spent by the particle under the barrier but
neither of them has been shown to be equivalent to the primary
notion of the tunneling time operationally defined by an
experiment with clocks. The problem has recently been
surveyed by R.Y.Chiao \cite{Chiao98} (see also \cite{Hauge89},
\cite{Landauer94}).
 
The experimental investigation of the problem became possible in
the last decade thanks to the recognition that massive particles
may be replaced by photons. In the experiments with
photons periodic layers of alternately
high and low index media serve as optical tunnel barrier. Such
experiments, while of great importance in themselves, may shed
light on the tunneling of nonrelativistic massive particles as
well, because the mathematical form of the equations, governing
the tunneling process, are essentially the same for both
cases.
 
The great advantage of using photons consists in the possibility
of converting time intervals
into phase shifts. This trick, going back to the Michelson-Morley
experiment, makes the measurement of exceedingly small
transit times feasible by the shift of the interference pattern in
an appropriate interferometer. In the Berkeley-experiments
\cite{Steinberg93}, \cite{Steinberg95} UV
photons were split into a pair of photons of equal frequency by
spontaneous frequency down conversion in a nonlinear medium. In a
Hong-Ou-Mandel spectrometer the two photons produce characteristic
minimum in their rate of coincidence when pass through a
beam-splitter which is in symmetrical position with respect to
the beams.
When, however, a thin tunneling layer was posed into the path of
one of the beams the coincidence minimum could be maintained only
if the beam-splitter was displaced at a certain distance. From
this distance the tunneling time of the photon could be deduced
and
turned out to be about 2 fs while the time required to traverse
the same distance in vacuum is 3.6 fs.
 
This result which is in rough agreement with the Wigner tunneling
time indicates that the speed of tunneling may exceed the speed of
light in vacuo by about 70 percent. It may be noted, however, that
stationary interference experiments, contrary to the experiments
with clocks, permit us to infer tunneling time 
only at the expense of imagining photons, moving in the arms of the
equipment, as {\em particles}.
It is this picture which might suggest the
counterfactual conditional statement that if
tunneling time was directly measured it would be equal to the time
inferred from the stationary interference experiment.
In quantum physics, however,
counterfactual statements based on imagined
rather then real events are in general invalid even if their validity in
classical physics would be beyond doubt.

This difficulty has been clearly recognized in \cite{Chiao98}, where it was
pointed out, that different experiments may lead to different tunneling
times. Based on this fact the
suggestion was made to abandon the unique definition of the tunneling time
and replace it by a multitude of
equivalent notions, assigning
to each conceivable experimental setup its own value of this quantity.
In particular, tunneling times measured in an experiment with clocks and
deduced from a particular stationary experiment are both legitimately called
tunneling time albeit under different circumstances.

But relativization of concepts often leads to confusion.
It seems, therefore, safer to keep
consistently to the definite meaning of tunneling time as
formulated at the
beginning of this section, because it conforms with the general
use
of the word "time", including its operational meaning. In doing so
one naturally has to admit, that the time parameters inferred from
experiments of other types are in general different from what is
properly called tunneling time.
This point of view may, perhaps, be argued for by noticing that in
relativity theory only such velocities are of significance which may be
related to real pairs of events through the elementary formula
$\Delta s/\Delta t$ 
while velocities inferred from stationary experiments are necessarily based on imagined
rather than real events. It may indeed be dangerously misleading 
to call "velocity" a quantity which lacks essential
connotations of this term.
 
The purpose of the present work is to suggest a theoretical scheme
to calculate the result of an experiment with clocks. The main
obstacle on the way to formulate such a method is the lack of a
rigorous
quantum theory of the arrival time distributions of photons and
particles. In quantum optics this difficulty has been overcome by
the replacement of the statement {\em "The detector has clicked"} with
the statement {\em "The detector atom is in one of its excited states"}.
Based on the assumed equivalence of these two statements
(see Section 8) working
formulas, known as counting rate formulas, were derived to
first order in the photon-detector interaction \cite{Glauber64}
which have since been used succesfully in treating space-time
correlations between photons.
 
The direct measurement of the tunneling time of photons is
actually the measurement of the time correlation between the
emission and the absorption of the photon on different sides of
the barrier. If, therefore, one accepts the basic assumptions
underlying counting rate formulas, an appropriate formula of this
kind can be derived which permits us to calculate
the value (or
distribution) of the tunneling time as operationally defined.
 
In the present work we confine ourselves (1) to the presentation
of this formula and (2) to show that in the absence of any barrier it
leads to the expected time correlation. In addition, it will be
illustrated
on a simplified model, how the theory 
works when a barrier is present. No
attempt is made to apply the method to realistic barriers as e.g.
to that used in the Berkeley-experiment. 

For a very high and broad barrier our model calculation gives
sharp tunneling time which is equal to zero and so it
agrees {\em qualitatively} with the result of
the Berkeley-experiment which predicts the reduction of the time
required to travel a given distance when a barrier is present. This
preliminary result may be an indication that the time parameter inferred
from the Berkeley-experiment is indeed closely related to the
tunneling time as defined in an experiment with clocks.

\section{Working formula for a thought experiment with clocks}
The experimental setup is shown on Fig. 1.
The photon source will be a two-level atom at the origin of the
coordinate system. An atomic detector at the point $\vec r =
(0,0,z)$ of
the $z$ coordinate axis serves to detect the photon. The barrier,
an infinite
homogeneous layer, supporting evanscent waves in
a broad interval around the photon wavelength, is placed between
the source atom and the detector perpendicular to the z-axis
within the region $(a,b)$ of this axis. An appropriate source
detector is assumed to be present in the immediate vicinity of
the source atom which clicks at the moment of the photon emission. It
may be assumed that it detects a particle 
(or another photon) which accompanies
instantaneously the emission of the tunneling photon. In short, the setup
works on the principle of the usual time of flight spectrometers.

\begin{center}
\epsfig{file=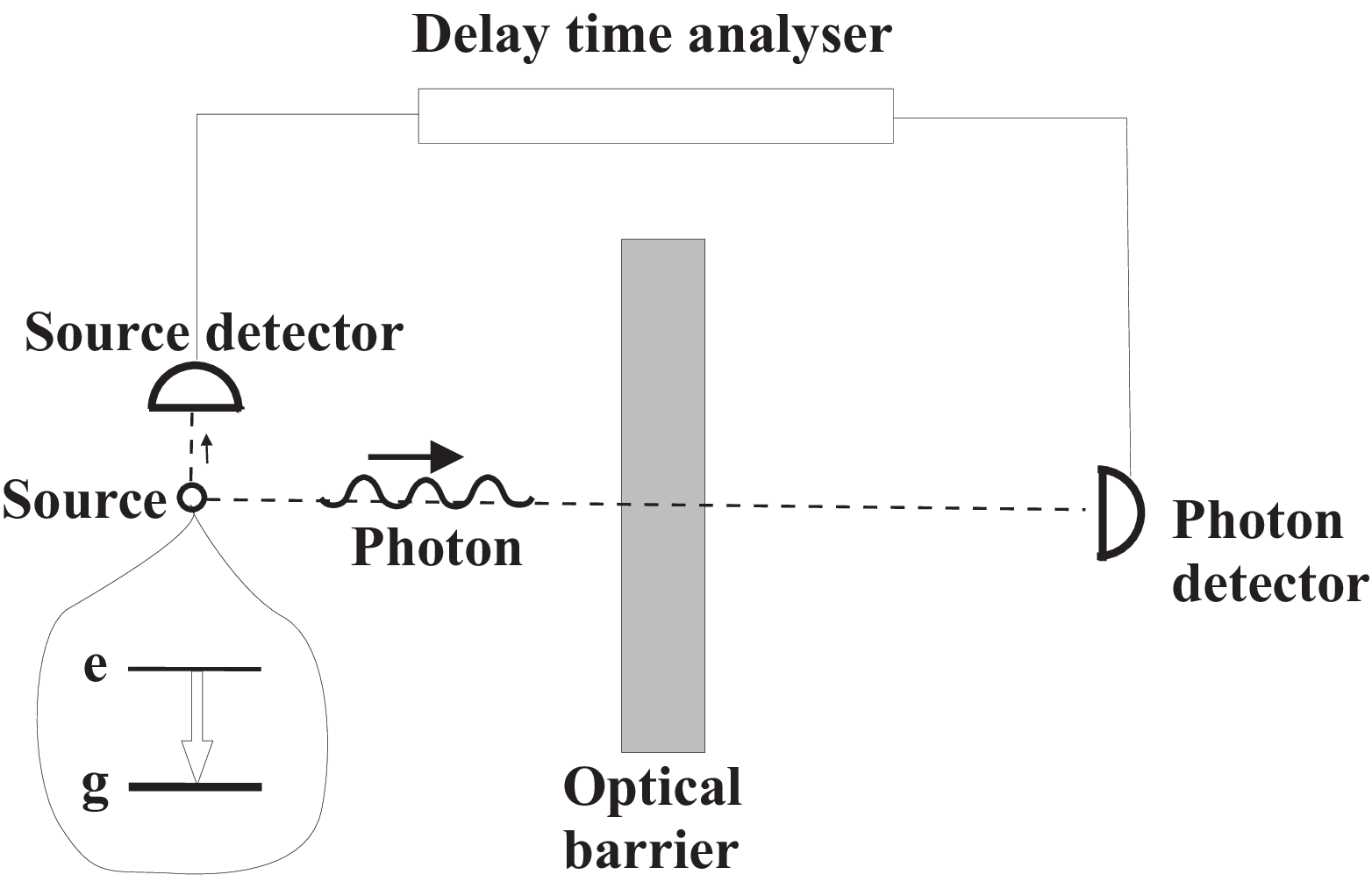,width=80mm}\\[2mm]
Figure 1.
\end{center}

\vspace{2mm}
 
Assume that at $t=0$ the source atom and the detector atom are in
their excited and ground states respectively and no photon is
present. Our aim is to calculate the probability density
$w(t_1,t_2)$ of the source detector clicking at $t_1$ and
photon detector clicking at $t_2$.
 
In order to calculate $w(t_1,t_2)$ one has first to determine the
probability $p(t_1,t_2)$ of finding the source in its
ground state at the moment $t_1$ and the photon detector
in some of its excited
states at $t_2$. Both $t_1$ and $t_2$ are chosen arbitrarily "by
ourselves" rather than by the experimental setup itself.
Then, having $p(t_1,t_2)$ calculated, $w(t_1,t_2)$ is obtained by
differentiation:
\begin{equation}
w(t_1,t_2) = \frac{\partial^2p(t_1,t_2)}{\partial t_1\partial
t_2}.\label{4a}
\end{equation}
This kind of detour through $p(t_1,t_2)$ has been used since the
sixties for the calculation of photon space-time correlations in
light beams prepared in various ways. In first Born-approximation
the detector degrees of freedom can be easily eliminated and
$p(t_1,t_2)$ and its generalizations can be expressed in terms of
the expectation value of some product of field variables.
Relations of this type are called counting rate formulae
\cite{Glauber64}. The restriction to first Born-approximation
ensures the nonnegativity of $w(t_1,t_2)$.
 
This same procedure will be adapted below for the calculation of
the
time correlation in photon tunneling. The version of the counting
rate
formulae appropriate for the purposes of the present work is the
following:
\begin{equation}
p(t_1,t_2) = \int_0^\infty\;d\omega\cdot\tilde{\sigma}(\omega
)\cdot\vert M_\omega (t_1,t_2)\vert^2,\label{5a}
\end{equation}
where $\tilde{\sigma}(\omega )$ is the spectral sensitivity of the
photon detector and
\begin{equation}
M_\omega (t_1,t_2) = \int_0^{t_2}\;dt\cdot e^{\displaystyle i\omega
t}
\langle g, vac|T\bigl (\varphi^H(t,\vec r)P_g^H(t_1)\bigr )|e,
vac\rangle .\label{5b}
\end{equation}
The amplitude in the last integral is the matrix element between
the states $|g, vac\rangle\equiv |g\rangle\otimes |vac\rangle$ and
$|e, vac\rangle\equiv |e\rangle\otimes |vac\rangle$ of the photon
vacuum and the source in its ground and excited states. $P_g$ is
the projector to the ground state of the source atom, $\varphi$
is the field operator of the photon, assumed spinless, and $\vec
r$ is the position of the photon detector. The superscript $H$
indicates Heisenberg picture in which the dynamical
variables are driven by the Hamiltonian $H = H_s + H_f + H_{sf}
\equiv H_0 + H_{sf}$ which is the sum of the source, the field and
the source-field interaction Hamiltonians. The last term is
assumed to be of the simple form
\begin{equation}
H_{sf} = Q\cdot\varphi (\vec r=0).\label{5d}
\end{equation}
The operator $Q$ acts in the Hilbert space of the source, its
nonzero matrix elements being $Q_{eg}={Q}_{ge}^*$. The symbol $T$
means time ordering. The main steps of the derivation of
(\ref{5a}) and (\ref{5b}) are summarized in Appendix A.
\section{Quantization of the photon field}
The equation satisfied by the field operator $\varphi$ is
\begin{equation}
\frac{\partial^2\varphi}{\partial t^2} - \triangle\varphi +
V\varphi = 0\label{5c}
\end{equation}
(the system of units $c=\hbar =1$ is adopted). The last term
represents the barrier, confined to the region $(a,b)$ of
the z-axis between the origin and the position of the photon
detector at $z$. The width $D$ of the barrier is, therefore,
equal to $b-a$. For the
time being no homogeneity of the barrier in the $z$-direction will
be required.
 
The axial symmetry of the experimental setup suggests that
quantization be performed in cylindrical coordinates:
\begin{eqnarray}
\varphi (t,\vec r) &=&
\sum_{m=-\infty}^\infty\sum_{s=\pm}\int_0^\infty\;
\frac{d\omega_rd\omega_z}{\sqrt{2\omega}}
\left [
a_{ms}(\omega_r,\omega_z)U_m(\omega_r|r)v_s(\omega_z|z)e^{\displaystyle
im\varphi}e^{\displaystyle -i\omega t} + \right.\nonumber\\
&&\left.
+a_{ms}^+(\omega_r,\omega_z)U_m^*(\omega_r|r)v_s^*(\omega_z|z)
e^{\displaystyle
-im\varphi}e^{\displaystyle i\omega t}
\right ],
\label{6a}
\end{eqnarray}
where
\begin{equation}
\omega = \sqrt{\omega_r^2 + \omega_z^2},\label{6b}
\end{equation}
and
\begin{equation}
U_m(\omega_r|r) = \frac{\sqrt{\omega_r}}{2\pi}
J_m(\omega_rr).\label{6c}
\end{equation}
 
The functions $v_\pm(\omega_z|z)$ obey the equation
\begin{equation}
-\frac{d^2v_\pm}{dz^2}+V(z)v_\pm = \omega_z^2v_\pm .\label{6d}
\end{equation}
The index $\pm$ indicates the direction of the incoming wave:
\begin{equation}
v_+(\omega_z|z) =
      \begin{cases}
      e^{\displaystyle i\omega_z(z-a)} +
      Re^{\displaystyle - i\omega_z(z-a)}&\text{if $z<a$},\\
      Te^{\displaystyle i\omega_z(z-a)}&\text{if $z>b$},
      \end{cases}
      \label{6e}
\end{equation}
 
\begin{equation}
v_-(\omega_z|z) =
      \begin{cases}
      Te^{\displaystyle
     -i\omega_z(z-b)}&\text{if $z<a$},\\
      e^{\displaystyle -i\omega_z(z-b)} + R'e^{\displaystyle
      i\omega_z(z-b)}&\text{if $z>b$},
      \end{cases}
      \label{6f}
\end{equation}
where we have used the fact that the transmission coefficient is
independent of the incoming direction ($T'=T$, see Appendix B).
These functions obey the relation
\begin{displaymath}
\sum_s\int\;d\omega_z\cdot v_s(\omega_z|z)v_s^*(\omega|z') =
2\pi\cdot\delta (z - z').
\end{displaymath}
Equation (\ref{6d}) is of the form of a
Schr\"{o}dinger-equation
for a particle moving along the z-axis (\cite{Martin},  \cite{Olkh}).
Therefore, the
transmission and reflection coefficients $T$ and $R$ are
the same which are found in the textbooks on Quantum Mechanics.
One has only to notice that, as the
asymptotic forms (\ref{6e}), (\ref{6f}) indicate, $T$ and
$R$ belong
to a barrier of the given shape shifted to the origin (i.e. to
$a=0$, $b=D$).
\section{The Wigner-Weisskopf approximation}
In the next section the matrix element in (\ref{5b}) will be
calculated in Wigner-Weisskopf (WW) approximation
\cite{Weisskopf30a}, \cite{Weisskopf30b}.
This approximation scheme is based on two assumptions.
 
The first assumption is the confinement of the electromagnetic
interaction to the subspace spanned by the
vectors $|e,vac\rangle $ end $|g,1\; photon\rangle $. The
special form (\ref{5d}) of the interaction and the relation
$J_m(0)=\delta_{m0}$ allow us to consider $m=0$ photons only. We
have, therefore, the state vector of the form
\begin{equation}
|t\rangle  = c(t)|e,vac\rangle  +
\sum_{s=\pm}\int_0^\infty\;d\omega_rd\omega_z
A_{\omega_r\omega_z}^s(t) |g,\omega_r\omega_zs\rangle ,\label{7a}
\end{equation}
in the interection picture, which obeys the
Schr\"{o}dinger-equation
\begin{equation}
i\frac{\partial |t\rangle }{\partial t} = H_{sf}(t)|t\rangle ,
\qquad \left (H_{sf}(t) = e^{\displaystyle iH_0t}H_{sf}(0)
e^{\displaystyle
-iH_0t}\right ), \label{8a}
\end{equation}
and the the initial condition $|t=0\rangle =|e,vac\rangle $.
It follows then
that the coefficients in (\ref{7a}) satisfy the equations
\begin{equation}
i\dot c(t) =
\sum_{s=\pm}\int_0^\infty\;d\omega_rd\omega_z
A_{\omega_r\omega_z}^s(t)
e^{\displaystyle i(\Omega -\omega
)t}\langle e,vac|H_{sf}(0)|g,\omega_r\omega_s\rangle \label{8b}
\end{equation}
\begin{equation}
i\dot A_{\omega_r\omega_z}^s(t) =
c(t)e^{\displaystyle i(\omega - \Omega
)t}\langle g,\omega_r\omega_z|H_{sf}(0)|e,vac\rangle .\label{8c}
\end{equation}
In these equations $\Omega$ is the excitation energy  $E_e - E_g$
of the source, and $\omega$ is given by (\ref{6b}).
 
The second assumption of the WW-method consists in the Ansatz
$\displaystyle c(t) = e^{\displaystyle - \Gamma t/2}$.
From (\ref{8c}) we then have
\begin{eqnarray}
A_{\omega_r\omega_z}^s(t) &=&
-i\int_0^t\;dt'\cdot e^{\displaystyle i(\omega - \Omega + i\Gamma
/2)t'}\langle g,\omega_r\omega_ss|H_{sf}(0)|e,vac\rangle  =
\nonumber\\
&&=\frac{Q_{ge}}{2\pi
}\sqrt{\frac{\omega_r}{2\omega}}\;v_s^*(\omega_z|0)
\frac{1-e^{\displaystyle i(\omega - \Omega + i\Gamma /2)t}}{\omega
 - \Omega
+ i\Gamma /2}.
\label{8d}
\end{eqnarray}
Now the value of $\Gamma$ could be computed from (\ref{8b}) but
since it is irrelevant for the present work we will not pursue
this line any further.
 
We notice that the exponential Ansatz is invalid for very short
times of the order of $1/\Omega$. This Ansatz prescribes the
nondecay amplitude $\langle e,vac|t\rangle $ as $\displaystyle
e^{\displaystyle
-\Gamma t/2}$ whose
time derivative at $t=0$ is equal to $-\Gamma /2$. However,
according to (\ref{8a}), this time derivative actually
vanishes:
\[i\left [\frac{\partial \langle e,vac|t\rangle }{\partial t}
\right ]_{t=0} =
\langle e,vac|H_{sf}(0)|e,vac\rangle  = 0.\]
Owing to this shortcoming of the WW-approximation our subsequent
considerations must be confined to the inside region of the future
light cone of the state preparation event at $\vec r = 0, t=0$. To
meet this condition the relations $t_1,\;t_2 > z$ --- or, more
precisely, $(t_1-z)\Omega\gg 1$, $(t_2-z)\Omega\gg 1$ --- will be
assumed from the outset.
This restriction is, however, irrelevant for a real experiment unless
the latter is specially designed to investigate the outside region
of the light cone (see Section 8).
 
The interaction picture of the present
section is connected with the Heisenberg-picture employed in
(\ref{5b}) by means of a unitary operator $W(t,0)$:
\begin{equation}
{\cal O}(t) = W(t,0){\cal O}^H(t)W^+(t,0)\label{9a}
\end{equation}
(notice that the absence of an upper index on operators indicates
interaction picture). Then, for (\ref{7a}) we have
\begin{equation}
|t\rangle  = W(t,0)|e,vac\rangle .\label{9b}
\end{equation}
Moreover, as a consequence of the first assumption the
radiative corrections to $|g,vac\rangle $ must be neglected and we
have the stability condition
\begin{equation}
W(t,0)|g,vac\rangle  = |g,vac\rangle .\label{9c}
\end{equation}
\section{The formula for $M_\omega (t_1,t_2)$ in WW
approximation}
The formula (\ref{5b}) for the amplitude $M_\omega (t_1,t_2)$
can be rewritten in the form
\begin{equation}
\begin{gathered}
M_\omega (t_1,t_2) =
\int_0^{t_2}\;dt\cdot e^{\displaystyle i\omega
t}\langle g,vac|P_g^H(t_1)\varphi^H(t,\vec r)|e,vac\rangle
+\\
+ \theta(t_2-t_1)\int_{t_1}^{t_2}\;dt\cdot e^{\displaystyle
i\omega t}
\langle g,vac|\bigl [\varphi^H(t,\vec r),P_g^H(t_1)\bigr ]|e,
vac\rangle ,
\end{gathered}\label{10a}
\end{equation}
where $\theta$ is the step-function.
 
When $t_1>t_2$, the probability amplitude of finding the source
deexcited and the photon detector excited is given by the first
term of (\ref{10a}) alone. One may expect that this amplitude
must not
depend on $t_1$ since at the earlier moment $t_2$ when the photon
detector was found excited the source was sure to having been
already in its ground state. In WW-approximation, due to the
stability condition (\ref{9c}), this assertion
is indeed true since $t_1$ drops out of the integrand of the first
term:
\begin{gather*}
\langle g,vac|P_g^H(t_1)\varphi^H(t,\vec r)|e,vac\rangle  =\\
= \langle g,vac|W^+(t_1,0)P_g(t_1)W(t_1,0)W^+(t,0)\varphi
(t,\vec r)W(t,0)|e,vac\rangle  =\\
= \langle g,vac|P_g(t_1)W(t_1,t)\varphi (t,\vec r)|t\rangle  =
\langle g,vac|\varphi
(t,\vec r)|t\rangle .
\end{gather*}
In addition to the stability condition use was made of the fact
that, in the interaction picture, $P_g(t_1)$ leaves $|g,vac\rangle
$ invariant. We can, therefore, write
\begin{equation}
\begin{gathered}
{\cal M}_\omega (t_2)\equiv
\int_0^{t_2}\;dt\cdot e^{\displaystyle i\omega
t}\langle g,vac|P_g^H(t_1)
\varphi^H(t,\vec r)|e,vac\rangle  =\\
= \int_0^{t_2}\;dt\cdot e^{\displaystyle i\omega
t}\langle g,vac|\varphi
(t,\vec
r)|t\rangle .
\end{gathered}\label{11d}
\end{equation}
Using (\ref{7a}), the matrix element here can be cast into the
form
\begin{equation}
\langle g,vac|\varphi (t,\vec r)|t\rangle  =
\sum_{s=\pm}\int_0^\infty\;d\omega_rd\omega_z
A_{\omega_r\omega_z}^s(t)\langle g,vac|\varphi (t,\vec r)
|g,\omega_r\omega_zs\rangle .\label{11a}
\end{equation}
The function $A_{\omega_r\omega_z}^s(t)$ is given by (\ref{8d})
while the matrix element in the integrand can be calculated using
(\ref{6a}):
\begin{equation}
\langle g,vac|\varphi (t,\vec r)|g,\omega_r\omega_zs\rangle  =
\frac{1}{2\pi}\sqrt{\frac{\omega_r}{2\omega}}\;e^{\displaystyle -i\omega
t}v_s(\omega_z|z)\label{11b}
\end{equation}
$\bigl (\mbox{remember that\ }\vec r = (0,0,z)\bigr )$.
 
The sum over $s$ can now be performed:
\begin{gather*}
\sum_{s=\pm}v_s^*(\omega_z|0)v_s(\omega_z|z) =
\left (e^{\displaystyle i\omega_za} + R^*e^{\displaystyle -
i\omega_za}\right )Te^{\displaystyle
i\omega_z(z-a)} +\\
+ T^*e^{\displaystyle -i\omega_zb}\left (e^{\displaystyle -
i\omega_z(z-b)} + R'e^{\displaystyle
i\omega_z(z-b)}\right ) =\\
= Te^{\displaystyle i\omega_zz} + T^*e^{\displaystyle -
i\omega_zz} + e^{\displaystyle
i\omega_zz}
\left (TR^*e^{\displaystyle -2i\omega_za} + T^*R'e^{\displaystyle
-2i\omega_zb}\right
).
\end{gather*}
According to (\ref{B1}) of the Appendix B the sum in the
parentheses is equal to zero:
\begin{equation}
TR^*e^{\displaystyle -2i\omega_za} + T^*R'e^{\displaystyle -2i\omega_zb}
 = 0,\label{B}
\end{equation}
therefore
\begin{equation}
\sum_{s=\pm}v_s^*(\omega_z|0)v_s(\omega_z|z) =
Te^{\displaystyle i\omega_zz} + T^*e^{\displaystyle -i\omega_zz}.
\label{11c}
\end{equation}
Putting now (\ref{8d}), (\ref{11b}) and (\ref{11c}) into
(\ref{11a})
and changing the integration variable from $\omega_r$ to $\omega =
\sqrt{\omega_r^2+\omega_z^2}$ we obtain
\begin{gather}
\langle g,vac|P_g^H(t_1)\varphi^H(t,\vec r)|e,vac\rangle  =
\langle g,vac|\varphi (t,\vec r)|t\rangle  =\notag \\
=
\frac{Q_{ge}}{8\pi^2}\int_0^\infty\;d\omega\cdot\frac{e^{\displaystyle
-i\omega t} - e^{\displaystyle -i(\Omega - i\Gamma /2)}}{\omega
- \Omega +
i\Gamma /2}\;\int_0^\omega\;d\omega_z\left [T(\omega_z)e^{\displaystyle
i\omega_zz} + T^*(\omega_z)e^{\displaystyle -i\omega_zz}\right
].\notag\\
\rule{0mm}{0mm}\label{12a}
\end{gather}
The matrix element in the second integral of (\ref{10a}) can be
handled analogously:
\begin{gather}
\langle g,vac|\bigl [\varphi^H(t,\vec r),P_g^H(t_1)\bigr ]|e,
vac\rangle  =\notag\\
=\sum_{s=\pm}\int_0^\infty\;d\omega_rd\omega_z
[A_{\omega_r\omega_z}^s(t_1)
- A_{\omega_r\omega_z}^s(t)]\langle g,vac|\varphi
(t,\vec r)|g,\omega_r\omega_zs\rangle  =\notag\\
=\frac{Q_{ge}}{8\pi^2}\left\{
e^{\displaystyle -i(\Omega - i\Gamma /2)t}
\int_0^\infty\;\frac{d\omega}{\omega - \Omega + i\Gamma /2}\;
\int_0^\omega\;d\omega_z\left [T(\omega_z)e^{\displaystyle i\omega_zz} +
T^*(\omega_z)e^{\displaystyle -i\omega_zz}\right ] -
\right.\notag\\
\left. -
e^{\displaystyle -i(\Omega - i\Gamma /2)t_1}
\int_0^\infty\;d\omega\frac{e^{\displaystyle -i\omega (t-t_1)}}{\omega -
\Omega + i\Gamma /2}\;
\int_0^\omega\;d\omega_z\left [T(\omega_z)e^{\displaystyle i\omega_zz} +
T^*(\omega_z)e^{\displaystyle -i\omega_zz}\right ] - \right
\}.\notag\\
\rule{0mm}{0mm}\label{12b}
\end{gather}
Here $t\ge t_1$ since this expression is the integrand in
the second term of (\ref{10a}).
 
Comparing (\ref{12b}) with (\ref{12a}) we see that in
WW-approximation
\begin{equation}
\langle g,vac|P_g^H(t_1)\varphi^H(t,\vec r)|e,vac\rangle  =
\langle g,vac|\bigl [\varphi^H(t,\vec r),P_g^H(0)\bigr ]|e,
vac\rangle .\label{12c}
\end{equation}
\section{Time correlation when no barrier is present}
In this case $T=1$, the $\omega_z$ integral in (\ref{12b})
gives $2\sin\omega z/z$ and we have
\begin{equation}
\langle g,vac|\bigl [\varphi^H(t,\vec r),P_g^H(t_1)
\bigr ]|e, vac\rangle ^0 =
\frac{Q_{ge}}{8\pi^2}\bigl [I_1^0 + I_2^0 + I_3^0 +
I_4^0\bigr ],\label{12d}
\end{equation}
where the superscript $0$ indicates the absence of the barrier and
\begin{equation}
\begin{gathered}
I_1^0(z) = \frac{1}{iz}e^{\displaystyle -i(\Omega - i\Gamma
/2)t}\int_0^\infty\;d\omega\frac{e^{\displaystyle i\omega z}}{\omega -
\Omega + i\Gamma /2},\\
I_2^0(z) = -\frac{1}{iz}e^{\displaystyle -i(\Omega - i\Gamma
/2)t_1}\int_0^\infty\;d\omega\;\frac{e^{\displaystyle -i\omega
(t-z-t_1)}}{\omega - \Omega + i\Gamma /2},\\
I_3^0(z) = -\frac{1}{iz}e^{\displaystyle -i(\Omega - i\Gamma
/2)t}\int_0^\infty\;d\omega\frac{e^{\displaystyle -i\omega z}}{\omega -
\Omega + i\Gamma /2},\\
I_4^0(z) = \frac{1}{iz}e^{\displaystyle -i(\Omega - i\Gamma
/2)t_1}\int_0^\infty\;d\omega\;\frac{e^{\displaystyle -i\omega
(t+z-t_1)}}{\omega - \Omega + i\Gamma /2}.
\end{gathered}\label{13a}
\end{equation}
Consider $I_1^0$. The integration contour can be deformed upward
to contain the positive imaginary axis and the quarter of the
large circle at infinity. Since $\omega_z$ is positive, the
integrand is exponentially small on this part of the large circle
and gives no contribution. Moreover, the
integrand is regular in the upper half plane and no pole
contributions arise. We have, therefore,
\begin{displaymath}
I_1^0(z) = \frac{1}{iz}e^{\displaystyle -i(\Omega - i\Gamma
/2)t}\int_0^\infty\;\frac{e^{\displaystyle -\eta z}}{i\eta -
\Omega + i\Gamma /2}\;i\:d\eta .
\end{displaymath}
This integral can be expanded in terms of the inverse of the large
distance $z$,  the leading term being of the order of $1/\Omega
z$, and so
$I_1^0$ turns out to be of second order. Therefore, in
the leading (linear) order in $1/z$, $I_1^0$ must be neglected.
 
The same conclusion applies to the sum of $I_3^0$ and $I_4^0$ as
well. In both of these terms pole contributions arise which are
linear in $1/z$ but they drop out of the sum.
 
Consider now $I_2^0$. When $t-z-t_1>0$ the contour has to be
deformed downward and a pole term
\begin{equation}
I_2^0(z) = \frac{1}{iz}\theta(t-z-t_1)\cdot 2\pi i\cdot e^{\displaystyle
-i(\Omega - i\Gamma /2)(t-z)}\label{14a}
\end{equation}
arises. In the opposite case of $t-z-t_1<0$ the deformation is
upward and no pole is to be dealt with. The contribution
of the integrals along the imaginary axis is negligible only if
$|t-z-t_1|\Omega\gg 1$. Therefore, $I_2^0(z)$ is given by
(\ref{14a})
provided the step function is assumed smoothed on the
scale $1/\Omega$. From an observational point of view such a
smoothing is of no significance and in what follows no attention
will be payed to it.
 
We have, therefore
\begin{equation}
\begin{gathered}
\langle g,vac|\bigl [\varphi^H(t,\vec r),P_g^H(t_1)\bigr ]|e,
vac\rangle ^0 =\\
=\frac{Q_{eg}}{4\pi z}\;\theta (t-t_1-z)e^{\displaystyle -
i(\Omega - i\Gamma /2)(t-z)}.
\end{gathered}\label{14b}
\end{equation}
From (\ref{12c}) we obtain
\begin{gather}
{\cal M}_\omega^0(t_2) =
-\int_0^{t_2}\;dt\cdot e^{\displaystyle i\omega t}
\langle g,vac|\bigl [\varphi^H(t,\vec r),P_g^H(0)\bigr ]|e,
vac\rangle ^0 =\notag\\
=i\frac{Q_{ge}}{4\pi z}\;\theta (t_2-z)e^{\displaystyle i(\Omega
- i\Gamma
/2)z}\;\frac{e^{\displaystyle i(\omega - \Omega + i\Gamma /2)t_2} -
 e^{\displaystyle
i(\omega - \Omega + i\Gamma /2)z}}{\omega - \Omega + i\Gamma
/2}.\notag\\
\rule{0mm}{0mm}\label{15a}
\end{gather}
The $\theta$-function here may in fact be omitted since $t_2>z$ by
assumption.
 
For the integral, occurring in the second term of (\ref{10a})
we obtain
\begin{gather}
\int_{t_1}^{t_2}\;dt\cdot e^{\displaystyle i\omega t}
\langle g,vac|\bigl [\varphi^H(t,\vec r),P_g^H(0)\bigr ]|e,
vac\rangle ^0 =\notag\\
=-i\frac{Q_{ge}}{4\pi z}\theta (t_2-t_1-z)e^{\displaystyle
i(\Omega - i\Gamma
/2)z}\;\frac{e^{\displaystyle i(\omega - \Omega + i\Gamma /2)t_2} -
 e^{\displaystyle
i(\omega - \Omega + i\Gamma /2)(t_1+z)}}{\omega - \Omega + i\Gamma
/2}.\notag\\
\rule{0mm}{0mm}\label{15b}
\end{gather}
Since $\theta (t_1) = \theta (t_2-z)=1$, the right hand side of
(\ref{15b}) is equal to
\begin{displaymath}
\theta (t_2-t_1-z)[{\cal M}_\omega^0(t_1+z)-{\cal
M}_\omega^0(t_2)].
\end{displaymath}
Hence, we have from (\ref{10a})
\begin{equation}
\begin{gathered}
M_\omega^0(t_1,t_2) = {\cal M}_\omega^0(t_2) +
\theta (t_2-t_1-z)[{\cal M}_\omega^0(t_1+z)-{\cal
M}_\omega^0(t_2)] =\\
=\theta (t_1 + z-t_2){\cal M}_\omega^0(t_2) +
\theta (t_2 - t_1 -z){\cal M}_\omega^0(t_1 + z).
\end{gathered}
\end{equation}
Substituting this into (\ref{5a}) we obtain
\begin{gather*}
p^0(t_1,t_2) = \theta (t_1+z-t_2)\int_0^\infty\;d\omega\cdot
\tilde{\sigma}(\omega )\vert{\cal M}_\omega^0(t_2)\vert^2 +\\
+\theta (t_2-t_1-z)\int_0^\infty\;d\omega\cdot\tilde{\sigma}
(\omega )\vert{\cal M}_\omega^0(t_1+z)\vert^2.
\end{gather*}
In arrival time measurements the spectral sensitivity must be as
broad as possible so we assume $\tilde{\sigma}(\omega ) =
\tilde{\sigma} = constant$. Then, substituting (\ref{15a}), we
find
\begin{equation}
\int_0^\infty\;d\omega\cdot
\tilde{\sigma}(\omega )\vert{\cal M}_\omega^0(t_2)\vert^2 =
\frac{\vert Q_{eg}\vert^2\tilde{\sigma}}{4\pi
z^2}\cdot\frac{1}{\Gamma}
\left (1 - e^{\displaystyle -\Gamma (t_2-z)}\right ),\label{16a}
\end{equation}
which leads to
\begin{gather}
p^0(t_1,t_2) =
\frac{\vert Q_{eg}\vert^2\tilde{\sigma}}{4\pi
z^2}\cdot\frac{1}{\Gamma}
\left \{1 - \theta (t_1+z-t_2)e^{\displaystyle -\Gamma (t_2-z)} - \theta
(t_2 - t_1 -z)e^{\displaystyle -\Gamma t_1}\right\}.\notag\\
\rule{0mm}{0mm}\label{16b}
\end{gather}
The formula (\ref{4a}) indicates that $w^0(t_1,t_2)$ may be
different from zero only around $t_1 + z - t_2 = 0$. Putting
\begin{displaymath}
w^0(t_1,t_2) = W(t_2)\cdot\delta (t_2 - t_1 - z),
\end{displaymath}
we find
\begin{gather*}
W(t_2) =
\int_{t_2-z-\epsilon}^{t_2-z+\epsilon}\;dt_1\cdot w^0(t_1,t_2) =
\int_{t_2-z-\epsilon}^{t_2-z+\epsilon}\;dt_1\cdot
\frac{\partial^2p^0(t_1,t_2)}{\partial t_1\partial t_2} =\\
=\left [\frac{\partial p^0(t_1,t_2)}{\partial t_2}\right
]_{t_1=t_2-z+\epsilon} -
\left [\frac{\partial p^0(t_1,t_2)}{\partial t_2}\right
]_{t_1=t_2-z-\epsilon}.
\end{gather*}
By (\ref{16b}) the second term is zero while in the limit of
$\epsilon =0$ the first one is equal to $\displaystyle\frac{\vert
Q_{ge}\vert^2\tilde{\sigma}}{4\pi z^2}\cdot e^{\displaystyle -\Gamma
(t_2-z)}$. Hence finally
\begin{equation}
w^0(t_1,t_2) =
\frac{\vert
Q_{ge}\vert^2\tilde{\sigma}}{4\pi z^2}\cdot e^{\displaystyle -\Gamma
t_1}\cdot\delta (t_2 - t_1 - z).\label{17a}
\end{equation}
Though this is just the expected result it is far from being an
obvious consequence of the counting rate formula
(\ref{5a}), (\ref{5b}) and of the reasoning in Appendix A
which led to it.
\section{Time correlation in the presence of a rectangular barrier}
Let us choose in (\ref{6d}) a constant $V(z)$ equal to $\mu$ in
the interval $(a,b)$ of width $D$ and zero outside. Since
(\ref{6d}) is, up to constant coefficients, identical to the
nonrelativistic Schr\"{o}dinger-equation, the transmission
coefficient of this rectangular barrier can be taken over
from quantum mechanics:
\begin{equation}
T(\omega_z) = e^{\displaystyle -i\omega_zD}\cdot e^{\displaystyle
-D\sqrt{\mu^2-\omega_z^2}}\cdot\frac{\displaystyle 1 -
e^{\displaystyle 4i\alpha
(\omega_z)}}{1 - e^{\displaystyle 4i\alpha (\omega_z)}\cdot
e^{\displaystyle
-2D\sqrt{\mu^2 - \omega_z^2}}},\label{17b}
\end{equation}
in which
\begin{displaymath}
e^{\displaystyle 2i\alpha (\omega_z)} = \frac{\omega_z - i\sqrt{\mu^2 -
\omega_z^2}}{\omega_z + i\sqrt{\mu^2 - \omega_z^2}}.
\end{displaymath}
The function $T(\omega_z)$ has a cut along $\omega_z > \mu$. The
physical values are those on the upper edge of the cut where
$\sqrt{\mu^2 - \omega_z^2} = -i\sqrt{\omega_z^2 - \mu^2}$.
 
The right hand side of (\ref{12b}) can now be written as the
sum
\begin{displaymath}
\langle g,vac|\bigl [\varphi^H(t,\vec r),P_g^H(t_1)\bigr ]|e,
vac\rangle  =
\frac{Q_{ge}}{8\pi^2}\bigl [I_1 + I_2 + I_3 +
I_4\bigr ],
\end{displaymath}
where
\begin{gather}
I_1(z) = e^{\displaystyle -i(\Omega - i\Gamma
/2)t}\int_0^\infty\;\frac{d\omega}{\omega -
\Omega + i\Gamma /2}
\int_0^\omega\;d\omega_z\cdot T(\omega_z )e^{\displaystyle
i\omega_zz},\notag\\
I_2(z) = -e^{\displaystyle -i(\Omega -
i\Gamma
/2)t_1}\int_0^\infty\;d\omega\;\frac{e^{\displaystyle -i\omega
(t-t_1)}}{\omega - \Omega + i\Gamma /2}
\int_0^\omega\;d\omega_z\cdot T(\omega_z )e^{\displaystyle
i\omega_zz},\notag\\
I_3(z) = e^{\displaystyle -i(\Omega -
i\Gamma /2)t}\int_0^\infty\;\frac{d\omega}{\omega -
\Omega + i\Gamma /2}
\int_0^\omega\;d\omega_z\cdot T^*(\omega_z )e^{\displaystyle
-i\omega_zz},\notag\\
I_4(z) = -e^{\displaystyle -i(\Omega -
i\Gamma
/2)t_1}\int_0^\infty\;d\omega\;\frac{e^{\displaystyle -i\omega
(t-t_1)}}{\omega - \Omega + i\Gamma /2}
\int_0^\omega\;d\omega_z\cdot T^*(\omega_z )e^{\displaystyle
-i\omega_zz}.\notag\\
\rule{0mm}{0mm}\label{18a}
\end{gather}
We will argue below that the formula
(\ref{17a}), derived in the absence of a barrier, remains
to reasonable accuracy valid also for a sufficiently
high and broad rectangular barrier provided $z$ is replaced in it
by $z-D$.
 
The variable $z$ in the formulae (\ref{18a}) appears only in
the inner
integrals. If the first exponential factor of $T$ is separated:
$\displaystyle T(\omega_z) = e^{\displaystyle -i\omega_zD}{\cal
T}(\omega_z)$, then the
exponentials in these integrals become $\displaystyle
e^{\displaystyle\pm
i\omega_z(z-D)}$.
 
The essential contribution to the $I_j$-s must come from the region of
integration around $\Omega$. If $\mu\gg\Omega$ the exponent
$\displaystyle
e^{\displaystyle -2D\sqrt{\mu^2 - \omega_z^2}}$ in the denominator is very small in
this region. We expand the fraction in $T(\omega_z)$ in terms of this
small quantity and retain from the resulting asymptotic expansion the
first term only (i.e. we disregard the second term of the denominator).
Then,
\begin{equation}
{\cal T}(\omega_z) = e^{\displaystyle -D\sqrt{\mu^2-\omega_z^2}}\left
[1 - \left
(\frac{\omega_z - i\sqrt{\mu^2-\omega_z^2}}{\mu}\right )^4\right
].\label{19a}
\end{equation}
Consider the behaviour of the factor $\displaystyle e^{\displaystyle
-D\sqrt{\mu^2 -
\omega_z^2}}$ when $|\omega_z|\longrightarrow\infty$ along some
direction $\varphi$ on the upper half $\omega_z$-plane. Then the
value $\varphi =0$ corresponds to the upper edge of the cut and
approaching infinity we have $\sqrt{\mu^2 - \omega^2}\sim
-i|\omega_z|$. Therefore, for positiv $\varphi$
\begin{displaymath}
\sqrt{\mu^2 - \omega_z^2}\sim -i|\omega_z|e^{\displaystyle i\varphi} =
-i|\omega_z|\cdot\cos\varphi + |\omega_z|\cdot\sin\varphi ,
\end{displaymath}
so that $\displaystyle\left\vert e^{\displaystyle -D\sqrt{\mu^2 -
\omega_z^2}}\right\vert$ approaches zero exponentially when
$|\omega_z|\longrightarrow\infty$.
 
Consider the inner integral
\begin{displaymath}
F(\omega )\equiv\int_0^\omega\;d\omega_z\cdot {\cal T}(\omega_z)
e^{\displaystyle i\omega_z(z-D)}
\end{displaymath}
in (\ref{18a}). Since its integrand vanishes exponentially on
the
large circle in the upper half plane the integration contour can
be deformed in this direction into two semiinfinite straight lines
along the positive imaginary direction:
\begin{gather*}
F(\omega ) =
i\int_0^\infty\;d\eta\cdot {\cal T}(i\eta )\cdot e^{\displaystyle
-\eta(z-D)} -\\
-ie^{\displaystyle i\omega (z-D)}\int_0^\infty\;d\eta\cdot
{\cal T}(\omega + i\eta
)\cdot e^{\displaystyle -\eta(z-D)},
\end{gather*}
which, to first order in $1/(z-D)$ is equal to
\begin{displaymath}
F(\omega ) = \frac{i}{z-D}[{\cal T}(0) - e^{\displaystyle i\omega
(z-D)}{\cal T}(\omega )].
\end{displaymath}
But ${\cal T}(0) = 0$, hence
\begin{displaymath}
F(\omega ) = \frac{e^{\displaystyle i\omega (z-D)}}{i(z-D)}{\cal T}
(\omega ).
\end{displaymath}
Substituting this into (\ref{18a}) we obtain
\begin{eqnarray*}
I_1(z) &=& \frac{1}{i(z-D)}e^{\displaystyle -i(\Omega - i\Gamma
/2)t}\int_0^\infty\;d\omega\frac{{\cal T}(\omega )e^{\displaystyle
i\omega
(z-D)}}{\omega - \Omega + i\Gamma /2},\\
I_2(z) &=& -\frac{1}{i(z-D)}e^{\displaystyle -i(\Omega - i\Gamma
/2)t_1}\int_0^\infty\;d\omega\;\frac{{\cal T}(\omega )e^{\displaystyle
-i\omega
(t-z+D-t_1)}}{\omega - \Omega + i\Gamma /2},\\
I_3(z) &=& -\frac{1}{i(z-D)}e^{\displaystyle -i(\Omega - i\Gamma
/2)t}\int_0^\infty\;d\omega\frac{{{\cal T}}^*(\omega )e^{\displaystyle
-i\omega
(z-D)}}{\omega - \Omega + i\Gamma /2},\\
I_4(z) &=& \frac{1}{i(z-D)}e^{\displaystyle -i(\Omega - i\Gamma
/2)t_1}\int_0^\infty\;d\omega\;\frac{{{\cal T}}^*(\omega)e^{\displaystyle
-i\omega (t+z-D-t_1)}}{\omega - \Omega + i\Gamma /2}.
\end{eqnarray*}
The direction of deformation of the contours in these integrals
remain the same as it was in (\ref{13a}) since the factor
$\displaystyle
e^{\displaystyle -D\sqrt{\mu^2 - \omega_z^2}}$ in ${\cal T}(\omega)$ leaves
the behaviour of the integrand at infinity unchanged. We may,
therefore, write for the first two line of (\ref{18a})
\begin{displaymath}
I_j(z) = {\cal T}(\Omega - i\Gamma /2)\cdot I_j^0(z-D)
\end{displaymath}
and an analogous expression with ${{\cal T}}^*$ for the remaining
lines. Hence
\begin{equation}
w(t_1,t_2) =
\frac{\vert
Q_{ge}\cdot {\cal T}(\Omega - i\Gamma /2)\vert^2\tilde{\sigma}}{4\pi
(z-D)^2}\cdot
e^{\displaystyle -\Gamma t_1}\cdot\delta (t_2 - t_1 - z + D).\label{21a}
\end{equation}
Since this is a sharp distribution it can be interpreted in terms
of a tunneling time equal to zero: The velocity of tunneling is
infinitely large. In the context of the present work this
behaviour is in no conflict with the requirement that no
information be transmittable faster than light. The reason is that
the WW-approximation limits the validity of our calculation to
$t_1,t_2 > z$, i.e. to the inside of the light cone $L$ of the
source state preparation event, which was the last occasion when
the experimentalist had access to the source (see Fig.2).
When the lifetime $\tau$ of the source atom is much larger than $z/c$ (which may
be of the order of several nanoseconds) the detection events fall
predominantly within $L$, permitting thereby no faster than light
information transfer. It is this region which is covered by our
calculation. An improved treatment valid for $\tau\le z/c$ too 
(i.e. in both the vicinity of $L$ and outside it)
would certainly be of great interest\footnote{The limit
$\tau\longrightarrow 0$ would be of special importance since it is
closely related to
the situation when the photon is released at the moment of
pressing a "release button" by the experimentalist in any freely chosen
instant of time. When $\tau\le z/c$ the precise nature of the state
preparation event requires also closer examination.}.
\begin{center}
\epsfig{file=b.eps,width=80mm}\\[2mm]
Figure 2.
\end{center}

For a real barrier --- or
even for our model barrier in a better approximation --- the tunneling
velocity will probably have a finite value which is greater than the
velocity of light in vacuo. From the point of view
of relativity theory, however, the
point of demarcation is at $c$. Hence,
the conlusions drawn from Figure 2  
remain practically unchanged for any tunneling velocity larger than $c$.

\section{Final remarks}
Tunneling time measurements are of two very different kinds: Stationary
measurements in which no moments of time are identified at all
and experiments in which moments of time of certain real events
are determined by using some kind of clocks.

The
Berkeley-experiment discussed in Sec.1 is an ingenious example of
the first type. No experiment of the second type has so far been
performed since it would require precise measurement of extremely small
time intervals. Even the purely theoretical analysis of this latter kind of
experiments presents a challenge. The present work
is an attempt to predict the result of such an experiment.
Though the calculation performed is based on a version of the counting
rate formulas widely used in quantum optics
it cannot be considered completely satisfactory
\cite{Glauber64}.
Quantum theory provides unambiguous rules for the calculation of
the probability $p(t_1,t_2)$ from which the correlation function
$w(t_1,t_2)$ is obtained by differentiation. Though
the rules of quantum theory ensure the positivity of $p(t_1,t_2)$
they don't render it a nondecreasing function of its
arguments and so the procedure may end with a negative
probability density.
 
The origin of this "positivity problem" may be traced back to the
replacement of the spontaneous state reduction of the detectors 
--- a process which falls 
outside the scope of the Schr\"{o}dinger equation --- by a "naive reduction
hypothesis", consisting in the identification of the statement {\em "The
detector has clicked"} with the statement {\em "The detector atom is in one of
its excited states"}. This replacement, however, may be accepted only in the
limit of weak coupling between the field and the detector
when the latter probability is always a nondecreasing
function of time. In the general case the rules of quantum theory do
not exclude the possibility that this probability decreases in some
intervals of time,
while the very notion of the "detector" is
irreconcilable with such behaviour: For a detector the
probability of being excited must never decrease.
 
A possible solution of the positivity problem would be to take
into account in the dynamics of the detector atom the influence of
the equipment which it is built into. This would result in
introducing some element of irreversibility into the detector's
behaviour which might lead to a never decreasing excitation
probability. Theories with spontaneous reduction
(\cite{Ghirardi86},
\cite{Diosi89}) might be also of significance in this respect.
Since to first order in the detector-field interaction no
positivity problem arises,
it may, perhaps, be reasonably expected that in
the weak coupling limit the future detector theory will be
essentially reduced to
our atomic detectors treated in first order perturbation
theory on the basis of the naive reduction hypothesis.
 
We may hence conclude that an experiment of the second kind might
well contradict the theory in its present state even if the calculations themselves
are irreproachable, reflecting thereby our insufficient knowledge
of quantum physics and, perhaps, suggesting the direction toward its completion.
 
The situation with the experiments of the first kind is quite the
opposite. They belong to the domain of phenomena where the
applicability of quantum theory has already been abundantly
demonstrated. Therefore, their outcome can in principle be calculated in
advance and the corresponding time parameter be deduced from this
calculation: no contradiction with known principles is expected.
In particular, superluminal tunneling under such stationary
circumstances never contradicts special relativity since the tunneling
process is not accompanied by flow of
information, referring to moments of time. In an experiment
with clocks, on the other hand, superluminal tunneling would in general
contradict relativity theory. Since the theoretical analysis of this
experiment performed in the present work does not exclude completely the
possibility of superluminal information transfer
(see the end of the previous section) the
situation deserves careful consideration.

{\bf Acknowledgement}: The author is deeply indebted to G\'{a}bor
Hrask\'{o} without whose inspiring curiosity this research
would not have been pursued.

\appendix
 
\section{Derivation of the counting rate formula}

In this Appendix the derivation of the counting rate formula
(\ref{5a}), (\ref{5b}) based on the lecture [4] is outlined.
 
Consider the system, consisting of the source atom, the photon
field and the atomic photon detector. The detector which
signalizes the moment of the photon emission need not be
considered explicitely. The total Hamiltonian of this system is
${\cal H} = H_s + H_f + H_{sf} + H_d + H_{df}$ where
$H_s,\;H_f,\;H_d$ are the Hamiltonians of the source, the field
and the detector respectively while $H_{sf},\;H_{df}$ are the
corresponding interactions.
 
In order to incorporate into the calculation the irreversible
nature of the observations of the source and the
photon detector at the moments $t_1$ and $t_2$ we assume that at
these moments the corresponding interactions $H_{sf}$ and $H_{df}$
are switched off. This assumption will be referred to as the
"irreversibility hypothesis".
 
The ground state and the excited states of the detector will be
labelled by $\gamma$ and $\epsilon$. The initial state of the
system is
\begin{displaymath}
|0) = |0\rangle \otimes |\gamma\rangle  \equiv |e,vac,\gamma ).
\end{displaymath}
Let us work in the Heisenberg picture (labelled by the
superscript
$h$) in which the dynamical quantities are driven by ${\cal H}$.
After the moment of the first observation the state of the system
becomes
\begin{displaymath}
|intermediate) =
      \begin{cases}
      P_g^h(t_1)|0) &\text{if $t_1<t_2$},\\
      {\cal P}_\epsilon^h(t_2)|0) &\text{if $t_2<t_1$}.
      \end{cases}
\end{displaymath}
Here $P_g$ and ${\cal P}_\epsilon$ are projectors on the ground
state of the source and the excited state $\epsilon$ of the
detector.
 
After the second observation the state becomes
\begin{displaymath}
|t_1,t_2,\epsilon) =
      \begin{cases}
      {\cal P}_\epsilon^h(t_2)P_g^h(t_1)|0) &\text{if
      $t_1<t_2$},\\
      P_g^h(t_1){\cal P}_\epsilon^h(t_2)|0) &\text{if $t_2<t_1$},
      \end{cases}
\end{displaymath}
wich can also be written as
\begin{displaymath}
|t_1,t_2,\epsilon ) = T\bigl ({\cal
P}_\epsilon^h(t_2)P_g^h(t_1)\bigr )|0),
\end{displaymath}
where $T$ denotes time-ordering.
 
After having performed the observations the source is in its
ground state, the detector is in one of its excited states and no
photon is present. The state $|t_1,t_2,\epsilon )$ is, therefore,
identical to $|g,vac,\epsilon )$ except that its norm is smaller
than unity. The probability $p(t_1,t_2)$ introduced in Sec.2 is
equal to the square of this norm summed over $\epsilon$:
\begin{equation}
p(t_1,t_2) = \sum_\epsilon\left\vert (g,vac,\epsilon |T\bigl
({\cal P}_\epsilon^h(t_2)P_g^h(t_1)\bigr )|0)\right\vert^2.
\label{A1}
\end{equation}
Our aim now is to eliminate from this formula the explicite
reference to the photon detector (except its spectral
sensitivity).
 
Introduce the interaction picture labelled by $i$ by means of the
unitary operator
\begin{displaymath}
V(t,0) = e^{\displaystyle i{\cal H}_0^st}\cdot e^{\displaystyle -
i{\cal H}^st},
\end{displaymath}
where $s$ indicates Schr\"{o}dinger-picture and ${\cal H}_0^s =
{\cal H}^s - H_{df}^s$. In this picture the development of the
states is governed by $H_{df}^i(t) = V(t,0)H_{df}^h(t)V^+(t,0)$:
\begin{displaymath}
i\dot V(t,0) = H_{df}^i(t)V(t,0),
\end{displaymath}
the solution of which to first order in the detector-field
interaction is
\begin{equation}
V(t,0) = 1 - i\int_0^t\;d\tau\cdot H_{df}^i(\tau ).\label{A2}
\end{equation}
Since $\dot{\cal P}_\epsilon^i = i[{\cal H}_0^s,{\cal
P}_\epsilon^i]=0$ we have ${\cal P}_\epsilon^i(t) = {\cal
P}_\epsilon^s$ and hence
\begin{displaymath}
T\bigl ({\cal P}_\epsilon^h(t_2)P_g^h(t_1)\bigr ) =
      \begin{cases}
      V^+(t_2,0){\cal P}_\epsilon^sV(t_2,t_1)P_g^i(t_1)V(t_1,0)
      &\text{if $t_1<t_2$},\\
      V^+(t_1,0)P_g^i(t_1)V(t_1,t_2){\cal P}_\epsilon^sV(t_2,0)
      &\text{if $t_2<t_1$}.
      \end{cases}
\end{displaymath}
These expressions are to be calculated to first order, using
(\ref{A2}).
 
The first line ($t_1<t_2$) gives
\begin{gather*}
V^+(t_2,0){\cal P}_\epsilon^sP_g^i(t_1) -\\
-iV^+(t_2,0){\cal P}_\epsilon^s\left [\int_{t_1}^{t_2}\;dt\cdot
H_{df}^i(t)P_g^i(t_1) + \int_0^{t_1}\;dt\cdot
P_g^i(t_1)H_{df}^i(t)\right ] =\\
=V^+(t_2,0){\cal P}_\epsilon^sP_g^i(t_1) - iV^+(t_2,0){\cal
P}_\epsilon^s\int_0^{t_2}\;dt\cdot T\bigl
(H_{df}^i(t)P_g^i(t_1)\bigr ).
\end{gather*}
For the second line ($t_2<t_1$) we have analogously
\begin{gather*}
V^+(t_1,0)P_g^i(t_1){\cal P}_\epsilon^s -\\
-iV^+(t_1,0)\left [\int_{t_2}^{t_1}\;dt\cdot
P_g^i(t_1)H_{df}^i(t){\cal P}_\epsilon^s + \int_0^{t_2}\;dt\cdot
P_g^i(t_1){\cal P}_\epsilon^sH_{df}^i(t)\right ].
\end{gather*}
In the first integral of the last line $t>t_2$. Therefore, by the
irreversibility hypothesis we have $H_{df}(t)=0$ in it. Moreover,
in the remaining term ${\cal P}_\epsilon^s$ can be brought in
front of the integral since $P_g^i(t_1)$ does not depend on
$H_{df}^s$. We have, therefore,
\begin{equation}
T\bigl ({\cal P}_\epsilon^h(t_2)P_g^h(t_1)\bigr ) =
C(t_2,t_1) -iV^+(t_1,0){\cal P}_\epsilon^s
\int_0^{t_2}\;dt\cdot T\bigl (H_{df}^i(t)P_g^i(t_1)\bigr
),\label{A3}
\end{equation}
where
\begin{displaymath}
C(t_2,t_1) =
      \begin{cases}
      V^+(t_2,0)P_g^i(t_1){\cal P}_\epsilon^s&\text{if
      $t_1<t_2$},\\
      V^+(t_1,0)P_g^i(t_1){\cal P}_\epsilon^s
      &\text{if $t_2<t_1$}.
      \end{cases}
\end{displaymath}
When (\ref{A3}) is substituted into (\ref{A1}) the term
$C(t_2,t_1)$
gives no contribution since ${\cal P}_\epsilon^s|0)=0$:
\begin{displaymath}
p(t_1,t_2) = \sum_\epsilon\left\vert (g,vac,\epsilon
|V^+(t_1,0){\cal P}_\epsilon^s\int_0^{t_2}\;dt\cdot T\bigl
(H_{df}^i(t)P_g^i(t_1)\bigr )|0)\right\vert^2.
\end{displaymath}
Now, to first order in $H_{df}$ the operator $V^+$ must
be replaced by unity and since $(g,vac,\epsilon |{\cal
P}_\epsilon^s = (g,vac,\epsilon |$, we have
\begin{displaymath}
p(t_1,t_2) = \sum_\epsilon\left\vert (g,vac,\epsilon
|\int_0^{t_2}\;dt\cdot T\bigl
(H_{df}^i(t)P_g^i(t_1)\bigr )|0)\right\vert^2.
\end{displaymath}
 
Assume now that $H_{df} = q\cdot\varphi$ where $q$ acts in the
Hilbert-space of the photon detector. For an arbitrary dynamical
quantity ${\cal O}$ the
$i$-picture and the Schr\"{o}dinger-picture are connected by the
relation
\begin{equation}
{\cal O}^i(t) =
e^{\displaystyle i{\cal H}_0^st}e^{\displaystyle
-i{\cal H}^st}{\cal O}^h(t)
e^{\displaystyle i{\cal H}^st}e^{\displaystyle -i{\cal H}_0^st}=
e^{\displaystyle i{\cal
H}_0^st} {\cal O}^se^{\displaystyle -i{\cal H}_0^st}.\label{A4}
\end{equation}
The Hamiltonian ${\cal H}_0$ is the sum of $H_d$ and the
Hamiltonian of the source-field system  $(H_s + H_f + H_{sf})$
which
commute with each other. Hence for $q$ equation (\ref{A4}) reduces
to
\begin{displaymath}
q^i(t) = e^{\displaystyle iH_d^st}qe^{\displaystyle -iH_d^st},
\end{displaymath}
and for such operators as $\varphi$ and $P_g$ which are
independent of the photon detector it gives ${\cal O}^i={\cal
O}^H$ where the superscript $H$ refers to the Heisenberg-picture
introduced in Sec.2.
 
Assuming, that $H_d^s|\gamma \rangle =0$ and $H_d^s|\epsilon
\rangle =\omega_\epsilon|\epsilon \rangle $ we have
\begin{gather}
p(t_1,t_2) = \sum_\epsilon
\vert\langle \epsilon |q|\gamma \rangle \vert^2\cdot
\left\vert \langle g,vac
|\int_0^{t_2}\;dt\cdot e^{\displaystyle i\omega_\epsilon t} T\bigl
(\varphi^H(t,\vec r)P_g^H(t_1)\bigr )|0,vac\rangle
\right\vert^2.\notag\\
\rule{0mm}{0mm}\label{A5}
\end{gather}
Since the spectral sensitivity is given by the relation
\begin{displaymath}
\tilde{\sigma }(\omega ) = \sum_\epsilon\delta (\omega -
\omega_\epsilon )\vert \langle \epsilon |q|\gamma \rangle \vert^2,
\end{displaymath}
(\ref{A5}) becomes identical to the working formulae given in
Sec.2.

\section{Derivation of the formula (\protect\ref{B})}

Consider the solution $v_+(\omega_z|z)$ of the equation
(\ref{6d})
given in (\ref{6e}). Since the equation is a real linear one,
the combination
\begin{eqnarray*}
&&v_+^*(\omega_z|z) - R^*v_+(\omega_z|z) =\\
&&=T^*e^{\displaystyle i\omega_z(a-b)}
      \begin{cases}
      \frac{\displaystyle 1-\vert R\vert^2}{\displaystyle
      T^*}\;e^{\displaystyle
      -i\omega_z(z-t)}&\text{if $z<a$},\\
      e^{\displaystyle -i\omega_z(z-b)}-
      \frac{\displaystyle R^*T}{\displaystyle
      T^*}\;e^{\displaystyle
      2i\omega_z(b-a)}e^{\displaystyle i\omega_z(z-b)}&\text{if
      $z>b$}.
      \end{cases}
\end{eqnarray*}
is also a solution which contains an incoming wave from the right.
Comparing this solution with (\ref{6f}) we have
\begin{eqnarray*}
T'&=& \frac{1-\vert R\vert^2}{T^*}\\
R'&=& - R^*\frac{T}{T^*}e^{\displaystyle 2i\omega_z(b-a)}.
\end{eqnarray*}
The first of these equations combined with the conservation of the
probability $\vert T\vert^2 + \vert R\vert^2 = 1$ gives $T' = T$
while the second one can be rewritten in the form
\begin{equation}
TR^*e^{\displaystyle -2i\omega_za} + T^*R'e^{\displaystyle -2i\omega_zb}
 = 0\label{B1}
\end{equation}
which is used in Sec.5.


\end{document}